\documentclass[graybox, natbib, footinfo]{svmult}

\usepackage{mathptmx}       % selects Times Roman as basic font
\usepackage{helvet}         % selects Helvetica as sans-serif font
\usepackage{courier}        % selects Courier as typewriter font
\usepackage{type1cm}        % activate if the above 3 fonts are
\usepackage{makeidx}         % allows index generation
\usepackage{graphicx}        % standard LaTeX graphics tool
\usepackage{multicol}        % used for the two-column index
\usepackage[bottom]{footmisc}% places footnotes at page bottom

\makeindex             % used for the subject index
\begin{document}
\title*{Photospheric constraints, current uncertainties in models of stellar atmospheres, and spectroscopic surveys}
\titlerunning{Photospheric constraints: current uncertainties}
\author{Bertrand Plez and Nicolas Grevesse}
\institute{Bertrand Plez \at Laboratoire Univers et Particules de Montpellier, CNRS, UniversitŽ\'e Montpellier 2, Montpellier, France, \email{bertrand.plez@univ-montp2.fr}
\and Nicolas Grevesse \at Centre Spatial de Li\`ege,  Universit\'e de Li\`ege, avenue Pr\'e Aily, B-4031 Angleur-Li\`ege,  Belgium \and Institut d'Astrophysique et de G\'eophysique, Université de Liège, All\'ee du 6 Ao\^ut, 17, B-4000 Li\`ege, Belgium, \email{nicolas.grevesse@ulg.ac.be}}
\maketitle

\abstract{We summarize here the discussions around photospheric constraints, current uncertainties in models of stellar atmospheres, and reports on ongoing spectroscopic surveys. Rather than a panorama of the state of the art, we chose to present a list of open questions that should be investigated in order to improve future analyses.}

\section{Introduction}
\label{sec:1}
We summarize here the questions that were raised during the discussion session following the talks on {\sl Photospheric constraints, current uncertainties in models of stellar atmospheres, and reports on ongoing spectroscopic surveys}.  Many of the issues that were discussed could not be settled, and most are also addressed in this volume's contributions. We refer to them in the text. This paper remains therefore mainly as a collection of questions, and tracks to be explored, in order to provide us with more definite answers on which we can rely for future work. 

\section{Model atmospheres}
It was shown by M. Asplund that present 3D models outperform all 1D models, despite their coarser description of the radiation field. Solar abundances derived with 3D models, all indicators, atomic and molecular, being consistent, are at present in conflict with helioseismology. The predicted sound speed, the depth of the convection zone, and the He/H surface abundance, all are in error. The solution of this problem will probably affect all stars. This is is a burning question, and its solution, either through a modification of the opacities, or some other mechanism is eagerly awaited. Could the solar 3D surface abundances increase again with further refinements of the models? This is unlikely, but a surprise is not to be excluded. Still, 3D models, obviously superior to 1D models, should be used in a broader range of studies, although the ease of use of 1D models make them hard to throw away from our toolboxes. A first approach is to use 1D averaged $\langle3\rm D\rangle$ models. They do not capture all the physics of full 3D atmospheres, but are as easy to use as a classical 1D model, in particular they can be used in NLTE studies. Grids should become available soon (M. Asplund).

\section{Abundances and stellar parameters}
T. Morel showed how the use of seismic gravities in 1D, LTE abundance analyses lead to discordances between neutral and ionized iron abundances. NLTE effects are of the order of 0.2 dex for dwarfs, and 0.4 dex for giants, the effect being greater at low metallicity, and lower gravity. Fe\,II lines seem to be a better iron abundance indicator, however, we know that iron lines are sensitive to 3D effects and to T$_{\rm eff}$. So far no 3D NLTE analysis of iron lines has been carried out. Reliable collisional cross sections with hydrogen and electrons are still missing for NLTE calculations. Progress will be made in the years to come that will probably improve the abundance determinations. For the time being we still don't know if Fe\,I or Fe\,II are reliable iron abundance indicators (T. Morel). 

It is important here to recall that spectroscopists will more and more use asteroseismic gravity (and T$_{\rm eff}$) determinations in order  to avoid using the iron ionization balance, or to test spectroscopic methods. Asteroseismologists on the other hand use spectroscopic gravities to test their inversion algorithms. We must be careful here not to go in circles, and be aware of the limitations on both sides. From what was quoted at this conference (e.g. D. Stello, and K. Brogaard) gravities from asteroseismology are more accurate than spectroscopic gravities, and this will remain true in the years to come. We must strive to understand the systematics behind this difference in gravity, using 3D and NLTE studies on representative samples of stars. The determinations of T$_{\rm eff}$ are not all in agreement. G. Zasowski showed the systematic differences between photometric and spectroscopic determinations. These must also be understood. Finally we should not forget uncertainties in atomic and molecular data (oscillator strengths, partition functions, ionization and dissociation potentials, etc).

\section{The contribution of large surveys}
The numerous on-going and planned surveys require a transition from detailed hand-made analyses of a few stars, using a few lines, to automated fits of very large samples using thousands of lines.
In the questions raised, the above detailed surveys will undoubtedly help. The large number of stars with parameters
spanning a large range of values helps highlight the errors, and delineate their causes (3D, NLTE, reddening, general calibration, ...). However, they must be completed by dedicated studies on smaller samples of well observed, well known objects, e.g. nearby stars with well-known distances, luminosities, diameters, masses. These detailed studies must be carried out with the best possible ensemble of techniques, models, and data. Current surveys do indeed include such studies in their design.

\section{Atmospheres as boundaries}
Model atmospheres can also, and are used as boundary conditions for stellar structure and evolution models. It was shown by P. Marigo that the surface properties of RGB models depend critically on that boundary condition and on the adopted convection theory. The effective temperatures can shift by almost 100\,K in response to a change of the mixing-length parameter of only 0.1. Detailed 3D model atmospheres, after proper 1D averaging, could be used to provide either a calibration of the mixing-length parameter, or better, thermal and pressure profiles (or whatever thermodynamical variable is appropriate) for inclusion in stellar evolution models. Maybe these 3D models should be calculated to deeper depths (M. Asplund). The same 3D models can be used to provide recipes for turbulent pressure and overshoot, or to better quantify mode excitation and damping rates (M. Asplund).

\section{Conclusion}
Confronting our results from asteroseismology and spectroscopy/photometry is an absolutely crucial step in our quest for better constrained stellar parameters, leading to our understanding of stellar and galactic evolution. Conferences like this one greatly contribute to making us all aware of the limitations of the various methods, encouraging us to join in our efforts to understand and resolve systematics and errors. 

%

%\input{referenc}
% BibTeX users please use
%\bibliographystyle{spphys}
%\bibliography{author_sesto}

% Mac users: please ignore the error message: "! Package natbib Error: Bibliography not compatible with author-year citations."
\end{document}